\documentclass[paper]{aa}     

\usepackage{graphics}
\usepackage{epsfig}
\usepackage{latexsym}
\begin{document}
\hyphenation{brems-strah-lung}
\title{INTEGRAL and Swift/XRT  observations \\ of the source PKS~0208-512}

\author{ Shu Zhang\inst{1}, Werner Collmar\inst{2}, Diego F. Torres\inst{3}, Jian-Min Wang\inst{1,4}, Michael Lang\inst{2}, Shuang-Nan Zhang\inst{1} 
}

\institute{Key Laboratory for Particle Astrophysics, Institute of High
Energy Physics, Beijing 100049, China
\and
 Max-Planck-Institut f\"ur Extraterrestrische Physik, P.O. Box 1603, 85740 Garching, Germany 
\and
ICREA \& Institut de Ci\`encies de l'Espai (IEEC-CSIC), Campus UAB, Facultat de Ci\`encies, Torre C5-parell, 2a planta, 08193 Barcelona, Spain 
\and
Theoretical Physics Center for Science Facilities (TPCSF), CAS
          }

\offprints{Shu Zhang}
\mail{szhang@mail.ihep.ac.cn}

\date{Received  / Accepted }

\titlerunning{INTEGRAL and Swift/XRT observations of the source PKS~0208-512}
\authorrunning{Shu Zhang et al.}

  \abstract
  {}
{The active galaxy PKS 0208-512, detected at lower energies by COMPTEL, 
has been claimed to be a MeV blazar from EGRET. We report on the most recent \emph{INTEGRAL} observations of the blazar PKS 0208-512,
 which are supplemented by \emph{Swift} ToO observations. }
  {The high energy X-ray and $\gamma$-ray emission of PKS 0208-512  during August - December  2008 has 
been studied using 682
ks of INTEGRAL  guest observer time and $\sim$ 56 ks  of \emph{Swift}/XRT  observations.
These data were collected during the decay of a 
$\gamma$-ray flare observed by \emph{Fermi}/LAT.}
{At X-ray energies (0.2 -- 10 keV) PKS 0208-512 is significantly detected by \emph{Swift}/XRT, showing 
a power-law spectrum with a photon index of $\sim$ 1.64. Its X-ray luminosity varied by roughly 30$\%$ during  one month. At hard X-/soft $\gamma$-ray energies PKS 0208-512 shows a marginally significant 
($\sim$ 3.2 $\sigma$) emission in the 0.5-1 MeV band when combining all \emph{INTEGRAL}/SPI data. 
Non-detections at energies below and above this band by \emph{INTEGRAL}/SPI may indicate intrinsic excess emission. 
If this possible excess is produced by the blazar, one possible explanation could be 
that its jet consists of an abundant electron-positron plasma, which   
may lead to the emission of an annihilation radiation feature. Assuming this scenario, we 
estimate physical parameters of the jet of PKS 0208-512. 
} 
{} 


   \maketitle
 
\section{Introduction}

It has been suggested that relativistic jets of Active Galactic Nuclei (AGN) may contain electron-positron pair plasmas
(Begelman et al. 1984). The detection of the high energy $\gamma$-ray emission from a very compact region around the
central supermassive black hole in M87 (Aharonian et al. 2006, Acciari et al. 2009) suggested
that electron-positron pair plasmas can be generated near black holes, thus setting up favorable conditions
for launching electron-positron pair plasma jets.  This is further supported by several AGN observations,
like those of PKS 2155-304 and Mkn 501, which have been detected to process much shorter (3-5 minutes) 
variability (Aharonian et al. 2007; Albert et al. 2007). 
Reynolds et al. (1996) argued for the dominance of an electron-positron pair plasma in the
jet of M87 and thus in other AGNs, and e.g.  Wardle et al. 1998, Hirotani et al. 2000, 
Lobanov \& Zensus 2001, Kino \& Takahara 2004, Dunn et al. 2006 studied different aspects of 
this eventuality. Other recent studies propose that 
extragalactic jets can also  be dynamically dominated by cold protons (e.g., Celotti \& Ghisellini 2008).
The determination of the jet content will certainly have important
implications for understanding the mechanisms of jet launching, acceleration and collimation.

The annihilation radiation in AGN jets may reveal itself in two different manners. Jets containing positrons hit a
dense ambient medium, then the positrons can be thermalized and annihilate with the ambient electrons, 
thereby forming a narrow peak around 511 keV. Marscher et al. (2007) searched for such a narrow
($\le$  3 keV) emission feature in the spectrum of the radio galaxy 3C 120, in which the jet 
strongly interacts with interstellar clouds. They did not find a signal, and their upper
limit did not constrain the positron-to-proton ratio in the jet. 
Alternatively, electrons and positrons may be already thermalized in the jets, and thus  
a broadened and blue-shifted annihilation component could show up on top of the spectral continuum 
in the MeV energy range (e.g., B\"ottcher \& Schlickeiser 1996; Skibo et al. 1997).
During the Compton Gamma Ray Observatory (\emph{CGRO}) era, several AGNs were reported to show such broad bumps at MeV energies 
(e.g., Bloemen et al. 1995; Blom et al. 1995), which were called  ``MeV blazars'' due to their excess emission at MeV energies. However, none of them passed subsequent data analysis checks 
(Blom et al. 1995; Williams et al. 2001; Stacy et al. 2003), and so these reports were discussed
quite controversially. In fact, the first convincing signal of electron-positron annihilation
radiation from AGN jets is yet to be found.

The source PKS 0208-512 is one such \emph{CGRO}-detected $\gamma$-ray source claimed to be a MeV blazar. Two
COMPTEL observations showed excess MeV-emission (Blom et al. 1995) compared to the extrapolation of the EGRET
spectrum, measured at energies above 100~MeV (Bertsch et al. 1993; Skibo et al. 1997).  
The variability of the source above 100 MeV was studied by von Montigny et al. (1995). They found
a characteristic variability timescale of eight days, which suggested a $\gamma$-ray emission region 
on the order of several Schwarzschild radii for a black hole of 10$^{10}$ M$_{\odot}$.
Further  $\gamma$-ray observations and data analyses
of this blazar led to the detection of persistent low-energy ($<$3~MeV) MeV emission at a significance level 
of about 4 $\sigma$ for the period 1991-1998 (Williams et al. 2001). A comparison to the contemporary
EGRET spectrum ($>$100 MeV) (Hartman et al. 1999) indicated a MeV excess. The strength of this excess however
is uncertain, because it varied with COMPTEL event selections (Williams et al. 2001). Furthermore, 
the statistical significance of this excess was challenged by analyses using all COMPTEL data (Stacy et al. 2003).
So, even today there is still  no unambiguous evidence for a MeV excess in the spectrum of PKS
0208-512, or any other blazar.

The satellite BeppoSAX observed PKS 0208-512 on 14th January 2001 and measured a spectral index of 
1.64$\pm$0.10 at soft X-rays (Donato et al. 2005). The Chandra observation 
(Schwartz 2006, 2007) with an exposure of $\sim$ 5 ks showed a complex X-ray, which can be resolved into at
least four regions: the dominant core, two regions along the jet that may correspond to a hot spot and 
an extended lobe, and a possible fourth region outside the jet. By adopting a CMB 
(cosmic microwave background) model, where the X-rays are assumed to be produced via Comptonization 
of the cosmic microwave background photons by jet blob electrons, the jet Doppler factor 
was estimated as 5.7-7.3. The fit of the Spectral Energy Distribution (SED), containing the 2008 \emph{Fermi} data at energies
above 200 MeV, resulted in a Lorentz factor of about 10 (Ghisellini et al. 2009).

We conducted \emph{INTEGRAL} observations of PKS 0208-512 between August and December 2008 for
a total of $\sim$ 682~ks. Since August 2008 the \emph{Fermi} observatory (Atwood et al. 2009) surveys the sky 
at energies above 30~MeV with its \emph{Large Area Telescope} (LAT), thereby also 
monitoring PKS~0208-512 on a daily basis. The source showed a flare in 2008. The long-term 
monitoring in optical and near-IR bands revealed a significant brightening of PKS~0208-512,
with an increase of 1.3 mag in the B-band and 1.4 mag in the R-band between September 11 and 30, 2008
(Buxton et al. 2008). A contemporaneous increase in $\gamma$-rays was detected by \emph{Fermi} 
(Tosti et al. 2008). The publicly available $\gamma$-ray lightcurve\footnote{at
http://fermi.gsfc.nasa.gov/ssc/data/access/lat/msl$\_$lc} shows a variable flux, up to about a factor of 6 on
timescales of weeks. Our \emph{INTEGRAL} observations  were carried out during the decay of this flaring event.
 Supplementary to our \emph{INTEGRAL}/SPI observations, we were granted two
contemporaneous \emph{Swift} ToO observations, providing the blazar state at softer X-ray energies. 
In this paper, we report on our findings.

\section{Observations and data analysis}

The ESA scientific mission \emph{INTEGRAL} is dedicated to high-resolution spectroscopy ($E/\Delta E \simeq 500$;
SPI see Vedrenne et al. 2003) and imaging (angular resolution: $12'$ FWHM, point source location accuracy:
$\simeq 1'-3'$; IBIS/ISGRI, see Ubertini et al. 2003 and Lebrun et al. 2003) of celestial sources in the energy
range $15$~keV to $10$~MeV. \emph{INTEGRAL} also provides simultaneous monitoring at X-rays ($3-35$~keV, angular
resolution: $3'$; JEM-X, see Lund et al. 2003) and optical wavelengths (Johnson V-filter, $550$~nm; OMC, see
Mas-Hesse et al. 2003). All \emph{INTEGRAL} instruments, except OMC, work with coded masks. Observations of
682 ks in total were carried out between August and December 2008 by applying the $5\times 5$ dithering mode
centered on PKS~0208-512. These data comprise 197 science windows (scw), each lasting typically for
3~ks.
 \emph{INTEGRAL}/SPI was operational for about 510 ks during these observations. Table~\ref{tab:integral-obs}
gives the details of the \emph{INTEGRAL} observations. The data are analyzed with the \emph{INTEGRAL}
\texttt{Offline Scientific Analysis (OSA) version 7.0}.  For the spectral analysis of the SPI data we used SPIMODFIT 
version 3.0, which is now also available in OSA version 8.0.

The $\gamma$-ray burst explorer \emph{Swift} was launched on November 20, 2004. It carries three co-aligned
detectors (Gehrels et al. 2004), namely the Burst Alert Telescope (BAT, Barthelmy et al. 2005), the X-Ray
Telescope (XRT, Burrows, et al. 2005) and the Ultraviolet/Optical  Telescope (UVOT, Roming et al. 2005).
Between 2005 and 2008 \emph{Swift}/XRT observed PKS~0208-512 13 times, exposing longer than 2 ks 
(Table~\ref{tab:swift-obs}). This yielded a total exposure time of $\sim$56 ks. The 2008
observations were -- by our ToO request -- contemporaneous to the \emph{INTEGRAL} observations. We selected all these
observations for our studies, and analyzed the XRT data with Heasoft v. 6.2. For the spectral fitting we applied
\texttt{XSPEC v 12.3.1} and estimated the model parameters at the $90$\% confidence level.

\section{Results}

\begin{figure}[ptbptbptb]
\centering
\includegraphics[angle=0, scale=0.38]{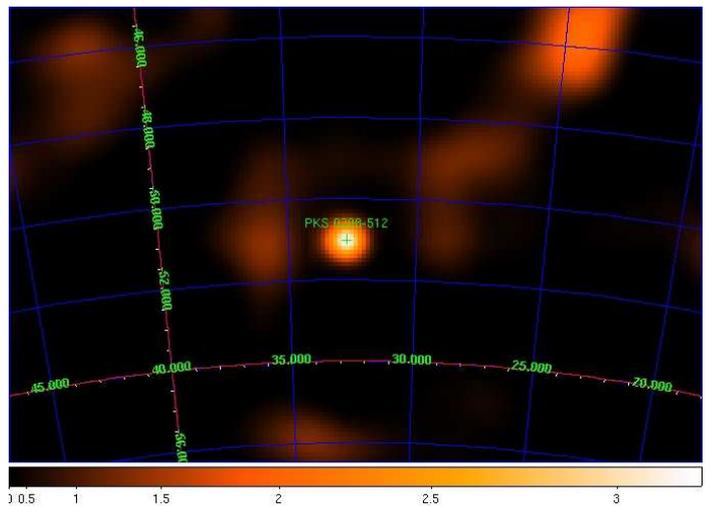}
\caption{
SPI significance map of the sky region centered on PKS 0208-512 in the 0.5--1 MeV  band
is shown for an exposure of $\sim$ 510 ks. A source at the position of PKS 0208-512, marked by a cross,
is clearly visible. The equatorial coordinates are overlaid, and the color bar
at the bottom gives the detection significance scale in $\sigma$.
}
\label{spimap}
\end{figure}

The source PKS~0208-512 was not detected by the \emph{INTEGRAL} instruments IBIS/ISGRI and JEM-X, neither in individual
scw nor in the sum of all data. However, the image of  \emph{INTEGRAL}/SPI, combining the whole data set ($\sim$
510 ks), reveals evidence for a detection at the 3.2 $\sigma$ level in the 0.5--1 MeV band
(Fig.~\ref{spimap}). 
A single source, exactly coincident in location with PKS~0208-512, is visible in the SPI map, 
which is otherwise rather empty and clean, at a flux level of
(1.50$\pm$0.47)$\times$10$^{-3}$ ph cm$^{-2}$ s$^{-1}$.  The source is not visible with SPI in 
other energy ranges, i.e. below 0.5 MeV and above 1 MeV. Maps in the 0.3--0.5 MeV band and above 1.0 MeV 
remain empty, i.e., yielding only upper flux limits.
 The SPI energy bands 0.3-0.5 MeV and 0.5-1.0 MeV adopted here are commonly used in SPI analyses. For example,  Kn\"odlseder et al. (2007) used 0.3-0.5 MeV and 0.514-1.0 MeV for producing SPI all-sky maps; Bouchet et al. (2005, 2008) used 0.3-0.6 MeV and 0.6-1.0 MeV bands, 
and Petry et al. (2009) chose 0.278-0.502 MeV and 0.502-1.0 MeV for imaging SPI sources.
We selected our bands before the analysis, based on these typical SPI choices.\footnote{ 
Above 1.0 MeV we took the high-energy band limit as 1.4 MeV because  the sensitivity above $\sim$ 1.4 MeV worsens
(see the SPI user manual, Dubath \& Kreykenbohm 2007).
We also notice that the presentation of the imaging results in an alternative binning, for instance using a constant ratio of the high to low energies defining the band, 
$E_{high}/E_{low}$=2, (i.e., bands as 0.25--0.5 MeV, 0.5--1.0 MeV, 1.0--2.0 MeV) differs little, if at all, from those reported in Table 1.}

In order to test for a possible flare or for a temporary instrumental effect, we also analyzed the SPI data
in sub-intervals. In each time interval, we found hints for an emission of the blazar, though at lower
statistical significance (Table~\ref{tab:integral-obs}), indicating a likely stable 0.5--1 MeV emission of
PKS~0208-512 across the SPI observations.

\begin{figure}[ptbptbptb]
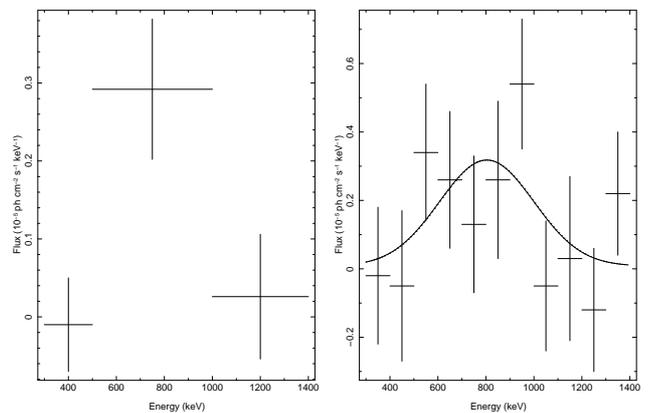

\centering
\includegraphics[angle=0, scale=0.23]{spe0-fit.ps}
\includegraphics[angle=0, scale=0.23]{spe-fit.ps}
\caption{
Broadband  \emph{INTEGRAL}/SPI spectrum from the cataloged sky position of PKS~0208-512 with the energy bins of the imaging analysis (left panel) and  a resolution of 100 keV (right panel).
The spectral analysis is consistent in behavior and flux with the results of the imaging analysis,
showing a weak ($\sim$ 3.2 $\sigma$) emission between 0.5 and 1 MeV, and no evidence for the source at energies below and above.  Over-plotted in the right panel is a Gaussian fit of the emission excess.
}
\label{spispc}
\end{figure}

 For the spectral analysis we used the cataloged sky position of PKS~0208-512 
and generated SPI spectra by running SPIMODFIT for narrow energy bands. The program SPIMODFIT was 
developed at the   Max-Planck-institut  f\"ur extraterrestrische Physik  (MPE) for the analysis of spectra of point sources. 
In order to test for possible instrumental effects, we generated the spectra for different event selections, i.e., using only
the SPI ``singles'' (events detected only in one SPI detector), using only the SPI ``multiples'' (events that
scattered once or twice inside SPI and so are ``seen'' by two or three detectors), and also using both event
types combined. In the first step we generated broadband spectra by similar energy cuts as in the imaging
analysis for the different event types. The spectra  of the ``single'' and ``multiple'' events are statistically
consistent with each other, and so independently indicate   a source flux
  at a level of $\sim$1.5$\times$10$^{-3}$ ph cm$^{-2}$ s$^{-1}$ in the 0.5 to 1 MeV band. 
Both spectra are also consistent in flux with the
 results of the imaging analysis.
Since both spectra show the same behavior, we then combined both event types for additional   analyses. The
resulting broadband spectrum is shown in Fig.~\ref{spispc} (left): 
It shows a flux level of (1.46$\pm$0.45)$\times$10$^{-3}$ ph
cm$^{-2}$ s$^{-1}$ in the 0.5--1 MeV band, and fluxes consistent with zero at energy bands above and below the latter. 
We repeated this analysis with a finer spectral binning to check for a possible narrow line emission. The
spectrum with a binning of 100~keV between 0.3 and 1.4 MeV is given in Fig.~\ref{spispc} (right).  It shows a
generally higher flux level in the 0.5 to 1.0 MeV band, which is consistent with the imaging analysis, 
but shows no obvious narrow band line emission. A subsequent analysis with an even finer binning of 50~keV resulted in the
same picture. The SPI spectrum (Fig.~\ref{spispc}) suggests a weak emission excess in the 
0.5-1 MeV band with respect to the neighboring energies. The width of this excess is constrained to
 $\sim$0.5~MeV by the SPI flux measurements at energies above and below the latter band. 
The results of the \emph{INTEGRAL} data analysis are given in Table~\ref{tab:integral-obs}.

The \emph{Swift}/XRT data are well represented by a simple power law during the time period between
2005 and 2008. The spectral index remained always at $\sim$ 1.6, but the flux dropped by $\sim$ 30$\%$
in the December 2008 observations compared to the previous ones. Table~\ref{tab:swift-obs} gives the details on
the \emph{Swift} observations and on the spectral results. 
The combined 2008 XRT spectrum is added to the SED
(Table~\ref{tab:swift-obs}, Fig.~\ref{sed})  to provide the contemporaneous X-ray spectrum.

\section{Discussion and summary}

In an exposure of $\sim$ 510 ks SPI found   hints at a $\sim$ 3.2-$\sigma$ level, for emission from PKS~0208-512
between 0.5 and 1 MeV at a flux of $\sim$1.5$\times$10$^{-3}$ ph cm$^{-2}$s$^{-1}$, without recognizable emission
at adjacent energies. 
 A $\chi^2$ test on this excess gives 1.1-$\sigma$ and  2.8-$\sigma$ deviations from a linear fit for the right and the left panels of the Fig. 2, respectively, consistent with  the low statistics of the flux. Far from being proven, the excess remains as an interesting possibility, whose consequences warrant analysis.

\begin{figure}[ptbptbptb]
\centering
\includegraphics[angle=0, scale=0.5]{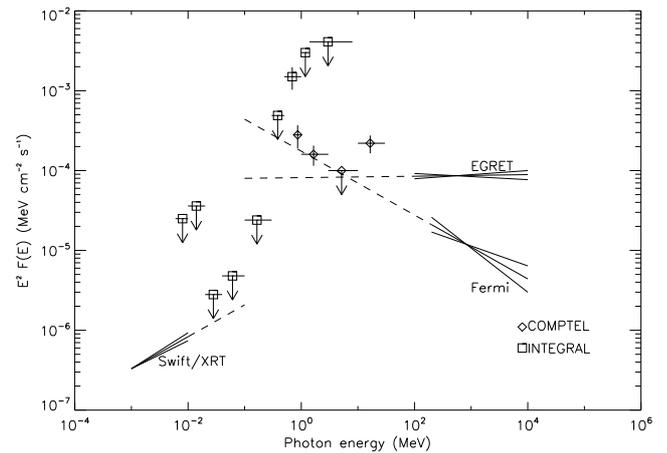}
\caption{
Contemporaneous high-energy SED of PKS 0208-512 for two different epochs,
the \emph{CGRO} era (COMPTEL, EGRET; 1991-2000) and recent 2008 measurements
(\emph{Swift}, \emph{INTEGRAL}, \emph{Fermi}). 
The COMPTEL data (open diamonds) are from Williams et al. (2001) covering
the time span 1991 to 1998, and the EGRET spectral shape is from Hartman et al. (1999),
measured between 1991 and 1994. The rest is from observations in 2008 by \emph{Swift}/XRT
(spectral shape; 0.2-10 keV), \emph{INTEGRAL} (JEM-X, IBIS/ISGRI, SPI; open squares;
6~keV-8~MeV), and \emph{Fermi}/LAT (spectral shape; above 200~MeV) between
 November 22 and December 27. The dashed lines
indicate the spectral extrapolations of the measured spectral shapes. The error bars
are 1 $\sigma$ and the upper limits are 2 $\sigma$.
}
\label{sed}
\end{figure}

To further investigate this possible emission, we analyzed the contemporaneous \emph{Fermi}/LAT data 
(collected during the SPI revolutions 746 to 757), and generated the simultaneous energy spectrum at energies 
above 200 MeV. This spectrum is added to the SED (Fig.~\ref{sed}). The direct extrapolation of this \emph{Fermi}/LAT spectrum to  soft $\gamma$-rays falls short compared to the measured SPI emission in the 0.5-1 MeV band. 
However, the unmeasured part of the high-energy spectra covers more than two decades, which may be misleading for a direct
comparison of the two measurements. 
Anyway, if we take a flux of $\sim$2$\times$10$^{-7}$ph cm$^{-2}$s$^{-1}$, as was measured contemporaneously by \emph{Fermi}/LAT, and connect it to the SPI flux in the 0.5--1 MeV band with a power-law shape, then the $\gamma$-ray photon index has to be steeper than three. On the other hand, the SPI upper limit at 0.3--0.5~MeV and the flux at 0.5--1 MeV require a photon index at hard X-ray energies  harder than 0.6. Therefore, if the measured SPI flux at 0.5--1 MeV is canonical inverse-Compton emission, the change in photon index from hard X-rays to $\gamma$-rays has to be larger than 2.4. This is hard to  account for in the current External Comptonization (EC) or Synchrotron Self-Comptonization (SSC) models, unless a very unusual electron energy spectrum is assumed. Consequently this might be indicative of an additional spectral component at 0.5--1 MeV in its SED. 

A similar trend of excess emission at soft $\gamma$-rays was indicated already about ten years ago, when COMPTEL and EGRET data were combined (e.g., Blom et al. 1995; Williams et al 2001). In Fig.~\ref{sed} we include the COMPTEL spectrum (Williams et al. 2001), where the flux at the lowest
COMPTEL energies (0.75-1 MeV) is about a factor of 3 higher than the extrapolation of the EGRET
spectrum. The results from
COMPTEL/EGRET and \emph{INTEGRAL}/SPI/\emph{Fermi} are independently derived and   may mutually strengthen each other. 

The broad emission feature between 0.5 and 1 MeV, if emitted by the blazar, could be understood as 
a broadened and Doppler blue-shifted pair annihilation radiation, emitted by a jet containing an 
electron-positron pair plasma. 
Assuming such a scenario, we derive by the arguments below the following estimates on the blazar jet 
of PKS~0208-512 and compare these to measured parameters using data from other wavelengths.

The
central energy $E$ of this emission feature is related to the kinematics of the jet,
$E=\gamma_{\rm min}D/(1+z) \times$ 511 keV, where $\gamma_{\rm min}$ is the minimum Lorentz factor of the
electrons and positrons in the jet, and $D$ is the Doppler boosting factor (B\"ottcher \& Schlickeiser 1996). 
 To estimate $E$, we fitted the right panel of Fig. 2 with a Gaussian shape. The central energy was derived as 803$^{+233}_{-291}$ keV, well within the energy range of 0.5--1.0 MeV, and the width was $\sim$ 200 keV without a well-constrained  error. The reduced $\chi^2$ for the fit was 1.1 for 7 degrees 
of freedom. 
The apparent speed of the relativistic bulk motion of PKS 0208-512 was measured with VLBI as $\sim$ (2.4$\pm$3.1)$c$ (Tingay et al. 2002). By taking $\gamma_{\rm min}=1$ and $D\sim$ 3, the  bulk Lorentz factor $\Gamma$ could then be estimated to  $\sim$ $2.6^{+3.0}_{-2.6}$.
Subsequently, the offset angle $\theta$ between our line of sight and the jet became $\sim$
10$^\circ$-19$^\circ$. We note that a caveat on the 
estimation of this offset angle is that the VLBI measurement of the bulk motion was not contemporary.  

 The density of a relativistic electron-positron pair plasma can be up to $\sim$
$10^{10}$$(L_{e^{\pm}}D^{-4}/V_{b})$$^{1/2}$ cm$^{-3}$ (Roland \& Hermsen 1995). Here the beam was assumed 
to consist of $e^{\pm}$ and the annihilation emission could dominate  in a volume $V_b$, which was inferred from the time variability in $\gamma$-rays. The relativistic $e^{\pm}$ in the beam were supposed to have a power law distribution with an index $\sim$ 3 (Roland \& Hermsen 1995). The annihilation luminosity could then be estimated under an approximation of the annihilation rate (Coppi \& Blandford 1990). If we take $L_{e^\pm}$ as the measured
luminosity at the 0.5--1 MeV band, $D$=3, and a lower limit of $V_{b}=\pi R_b^2L_b\sim5\times10^{46}$ cm$^3$, we  derive an
upper limit of $\sim$ $10^{9}$ cm$^{-3}$ for the density of electron-positron population in the jet. Here we
took the radius of the beam as $R_{b}$ $\sim$ $20 GM/c^2$ (Marcowith et al. 1995), 
the beam length $L_{b}$ of $\sim$ 100 
light days due to the  variability time scale of the  $\gamma$-rays flare observed by  \emph{Fermi}   in 2008,
and the central black hole mass as $M$ $\sim$ $10^8M_{\odot}$, which is typical for a blazar. The $L_{e^\pm}$ is about 6.3 $\times10^{48}$ erg/s, given a redshift $z$=1.003 (we used $H_0$=75 km s$^{-1}$Mpc$^{-1}$, $q_0$=0.5). By taking the Eddington limit of 1.3 $\times10^{38}$ M/M$_\odot$ erg/s, the central black hole mass was estimated to be larger than 6$\times$10$^{8}$ M$_\odot$. Using this mass, the upper limit for the density of the electron-positron population in the jet 
decreases to  $\sim$ $10^{8}$ cm$^{-3}$.  
A significant number of cold leptons, responsible for the annihilation line, may generate the  bulk spectral feature in the soft X-ray energy range via Comptonization off the surrounding UV photons, as discussed in e.g. Sikora et al. (1997), where the emission peak is estimated at $\sim$ $(\Gamma/10)^2$~keV. By taking $\Gamma$ $\sim$ 2.6, the emission peak is about 0.07 keV, well below the Swift/XRT energy domain. We notice that the XRT spectrum can be well fitted by a simple power-law shape, without any further components.

The gamma-ray flux above 100 MeV, averaged from August to October 2008, was a factor of $\sim$ 3.5 lower than
roughly ten years (1991-2000) ago during the \emph{CGRO} era (Hartman et al. 1999; Abdo et al. 2009). The \emph{Swift}/XRT observations show that the X-ray flux (2--10 keV) of PKS~0208-512 also decreased by more than 50\% in comparison
to the BeppoSAX observation in 2001 (Tavecchio et al. 2002; Donato et al. 2005). However, if the the soft $\gamma$-ray excess observed by SPI at 0.5-1 MeV is emitted by the blazar, this component is even brighter now than during the CGRO times.
A possible explanation could be that the
$\gamma$-rays generated by inverse-Compton emission of the non-thermal pairs are more dependent on the Doppler
factor than the thermal annihilation radiation in the jet (Skibo et al. 1997). Therefore, blazars, viewed
at a moderate jet offset-angle, can show a significant blue-shifted annihilation radiation,  which outshines the
continuum emission (Skibo et al. 1997). This may actually have happened in PKS 0208-512: a viewing angle of $\sim$
10$^\circ$-19$^\circ$ matches the prediction for this viewing-angle scenario. The moderate distance at a redshift
 $z\sim$1 makes PKS 0208-512 detectable in gamma-rays, despite its relatively large jet offset-angle.

Further monitoring of this source by \emph{INTEGRAL}, \emph{Fermi}/LAT and \emph{Swift}, 
as well as by the planned
NeXT high energy astronomy mission with a more sensitive Compton telescope below 600 keV and a broad band
capability down to about 0.3 keV should clarify this picture. If confirmed, the monitoring will be
able to constrain the pair density in the PKS~0208-512 jet directly from observing the annihilation
radiation of the pair plasma.

\acknowledgements
We are grateful to the anonymous referee for his/her comments which were of great helpful to polish the paper.
This work was subsidized by the National Natural Science Foundation of China, and the CAS key
Project KJCX2-YW-T03 and the 973 Program 2009CB824800. DFT has been supported by grants AYA2009-07391 and SGR2009- 811. J.-M. Wang and S.-N. Zhang thank the Natural Science Foundation of China for
support via NSFC-10325313, 10733010, 10725313 and 10821061.
Shu Zhang would like to thank \emph{INTEGRAL} and \emph{Swift} for approving  \emph{INTEGRAL} proposal No. 0620052
and two SWFIT ToO (Target of Opportunity) proposals  (ID No. 35002), and for subsequently carrying out the  observations of roughly 700 ks to support this research.

\begin{table*}
\fontsize{9pt}{10pt}\selectfont
\caption{
\emph{INTEGRAL} observations of PKS~0208-512. 
}

\centering

\begin{tabular}{cccccc}
\hline
\multicolumn{1}{c}{Rev. ID}&\multicolumn{1}{c}{Date}&\multicolumn{1}{c}{MJD}&\multicolumn{1}{c}{ SCWs}&\multicolumn{1}{c}{Exposure}&\multicolumn{1}{c}{Flux(0.5--1 MeV)}\\
\multicolumn{1}{c}{}&\multicolumn{1}{c}{ }&\multicolumn{1}{c}{(day)}&\multicolumn{1}{c}{ }&\multicolumn{1}{c}{(ks)}&\multicolumn{1}{c}{($\times$10$^{-3}$ ph cm$^{-2}$s$^{-1}$)}\\  \hline
714 & 2008 Aug 18-20 & 54696.26-54698.44 & 50  &172&\multicolumn{1}{c}{}\\
746 & 2008 Nov 22-23 & 54792.75-54793.75 & 24  &82&\multicolumn{1}{c}{1.5$\pm$1.2}\\
754 & 2008 Dec 16-17 & 54816.69-54817.77 & 25  &85& \multicolumn{1}{c}{1.1$\pm$1.1}\\
755 & 2008 Dec 19-21 & 54819.92-54820.28 & 31  &106& \multicolumn{1}{c}{2.5$\pm$1.0}\\
756 & 2008 Dec 22-23 & 54822.92-54823.74 & 19  &69& \multicolumn{1}{c}{1.6$\pm$1.3}\\
757 & 2008 Dec 24-26 & 54824.69-54826.78 & 48  &168&\multicolumn{1}{c}{ 1.2$\pm$0.8}\\  \hline
\hline
\multicolumn{1}{c}{Exposure}&\multicolumn{1}{c}{$E$-band}&\multicolumn{1}{c}{Flux }&\multicolumn{3}{c}{INTEGRAL}\\
\multicolumn{1}{c}{(ks)}&\multicolumn{1}{c}{(MeV)}&\multicolumn{1}{c}{($\times$10$^{-4}$ph cm$^{-2}$s$^{-1}$)}&\multicolumn{3}{c}{(instrument)}\\   \hline
510 &0.3--0.5&$<$6.4&\multicolumn{3}{c}{ }\\
& 0.5--1.0&15.0$\pm$4.7&\multicolumn{3}{c}{SPI(imaging)}\\
&1.0--1.4&$<$8.5&\multicolumn{3}{c}{}\\
&1.4--8.0&$<$24.2&\multicolumn{3}{c}{}\\ \hline
510 &0.3--0.5&$<$6.0&\multicolumn{3}{c}{}\\
& 0.5--1.0&14.6$\pm$4.5&\multicolumn{3}{c}{SPI(spectral)}\\
&1.0--1.4&$<$9.3&\multicolumn{3}{c}{}\\ \hline
682 &0.02--0.04&$<$0.7&\multicolumn{3}{c}{}\\
& 0.04--0.1&$<$0.7&\multicolumn{3}{c}{ISGRI}\\
&0.1--0.3&$<$1.6&\multicolumn{3}{c}{}\\\hline
682&0.006--0.01&$<$17.0&\multicolumn{3}{c}{JEM-X}\\
& 0.01--0.02&$<$17.6&\multicolumn{3}{c}{}\\\hline

\end{tabular}
\label{tab:integral-obs}
\begin{list}{}{}
\item[Note:]{The  revolution ID, calendar date,
time in MJD, number of science windows, exposure, energy band and flux are given. The errors are 1 $\sigma$ and upper limits 2 $\sigma$.}
\end{list}

\end{table*}

\begin{table*}
\fontsize{9pt}{10pt}\selectfont
\caption{\emph{Swift}/XRT observations   of PKS~0208-512. }
\centering
\begin{tabular}{ccccccccc}
\hline
\multicolumn{1}{c}{Obs. ID}&\multicolumn{1}{c}{Date}&\multicolumn{1}{c}{MJD }&\multicolumn{1}{c}{Offset}&\multicolumn{1}{c}{Expo.}&\multicolumn{1}{c}{Index}&\multicolumn{1}{c}{N}&\multicolumn{1}{c}{L}&\multicolumn{1}{c}{$\chi^2$/dof}\\
\multicolumn{1}{c}{(Swift/XRT)}&\multicolumn{1}{c}{ }&\multicolumn{1}{c}{(day)}&\multicolumn{1}{c}{(arcsecond)}&\multicolumn{1}{c}{(ks)}&\multicolumn{1}{c}{}&\multicolumn{1}{c}{}&\multicolumn{1}{c}{}&\multicolumn{1}{c}{}\\ \hline
00035002001 & 2005 Apr 23 & 53483.67   & 0.8& 12.4& & & & \\
00035002003 & 2005 May 04 & 53495.00   & 1.1& 2.1& & & & \\
00035002004 & 2005 May 12 & 53502.02   & 1.1& 2.1& & & & \\
00035002005 & 2005 May 10 & 53500.01   & 1.3& 4.2& & & & \\
combined    &            &         &    & 20.8&  1.64$\pm$0.06&4.4$\pm$0.2&3.4&0.9/46\\\hline
00035002014 & 2008 Oct 23 & 54762.56   & 1.5& 3.7& & & & \\
00035002017 & 2008 Nov 10 & 54780.65   & 2.3& 3.8& & & & \\
00035002021 & 2008 Nov 25 & 54795.43   & 1.5& 2.2& & & & \\
combined    &            &         &    & 9.7&  1.63$\pm$0.11&4.3$\pm$0.3&3.4&1.28/19\\\hline
00035002026 & 2008 Dec 17 & 54817.31   & 4.1& 4.2& & & & \\
00035002027 & 2008 Dec 20 & 54820.07   & 0.8& 4.2& & & & \\
00035002028 & 2008 Dec 20 & 54820.73   & 0.4& 4.4& & & & \\
00035002029 & 2008 Dec 23 & 54823.07   & 1.1& 4.4& & & & \\
00035002030 & 2008 Dec 24 & 54824.81   & 0.2& 4.2& & & & \\
00035002031 & 2008 Dec 26 & 54826.10   & 0.8& 4.6& & & & \\
combined    &            &         &    & 26&  1.64$\pm$0.07&3.0$\pm$0.2&2.3&1.2/38\\\hline
combined all   &            &         &    & 56.5&  1.62$\pm$0.04&3.9$\pm$0.1&3.1&1.25/100\\\hline
combined 2008  &            &         &    & 35.7&  1.60$\pm$0.05&3.3$\pm$0.1&2.7&1.1/59\\\hline
\hline
\end{tabular}
\label{tab:swift-obs}

\begin{list}{}{}
\item[Note:]{ The observation ID, time in calendar date and MJD,  the pointing offset angle,  exposure, spectral index, normalization (in units of 10$^{-4}$ ph cm$^{-2}$s$^{-1}$keV$^{-1}$), luminosity (in units of 10$^{-12}$ ergs cm$^{-2}$s$^{-1}$, at 0.2--10 keV), and the reduced $\chi^2$ and degree of freedom (dof) of the power-law fits are given. }
\end{list}
\end{table*}

\end{document}